\documentclass[titlepage,12pt]{article}

\usepackage{textcomp}
\usepackage{chetnew}
\usepackage{amssymb,amsmath,color,graphics,amscd,epsf,indentfirst,amsfonts}
\usepackage[utf8]{inputenc}
\usepackage{mathtools}
\usepackage{enumerate}
\usepackage{epsfig}
\usepackage{slashed}
\usepackage{tikz}

\def\blfootnote{\xdef\@thefnmark{}\@footnotetext}

\long\def\symbolfootnote[#1]#2{\begingroup%
\def\thefootnote{\fnsymbol{footnote}}\footnote[#1]{#2}\endgroup}

\makeatletter
\renewcommand{\@dotsep}{4.5}
\makeatother

\def\be{\begin{equation}}
\def\ee{\end{equation}}

\makeatletter
\def\@seccntformat#1{\csname the#1\endcsname.\quad}
\makeatother

\def\clock{{\count0=\time
           \divide\count0 60
           \ifnum\count0<10 0\fi\the\count0
           \multiply\count0 -60 \advance\count0 \time
           :\ifnum\count0<10 0\fi \the\count0
         }}
\newcommand{\timestamp}{{\small\vbox{\hbox{\tt\jobname.tex}
\hbox{\the\day/\the\month/\the\year, \clock}}}}

\def\time{{\boldsymbol{\tau}}}
\def\tt{{$tt^*$}}

\def\beq{\begin{equation}}
\def\eeq{\end{equation}}
\newcommand{\bea}{\begin{eqnarray}}
\newcommand{\eea}{\end{eqnarray}}
\def\bal{\begin{align}\begin{split}}
\newcommand{\eal}{\end{split}\end{align}}

\newcommand{\drawsquare}[2]{\hbox{%
\rule{#2pt}{#1pt}\hskip-#2pt
\rule{#1pt}{#2pt}\hskip-#1pt
\rule[#1pt]{#1pt}{#2pt}}\rule[#1pt]{#2pt}{#2pt}\hskip-#2pt
\rule{#2pt}{#1pt}}

\newcommand{\Yfund}{\raisebox{-.5pt}{\drawsquare{6.5}{0.4}}}
\newcommand{\Yasymm}{\raisebox
{-3.5pt}{\drawsquare{6.5}{0.4}}\hskip-6.9pt%
                      \raisebox{3pt}{\drawsquare{6.5}{0.4}}%
                     }
\newcommand{\Ysymm}{\Yfund\hskip-0.4pt%
                     \Yfund}

\newcommand\T{\rule{0pt}{2.5ex}}

\newcommand\B{\rule[-1.7ex]{0pt}{0pt}}

\def\drawbox#1#2{\hrule height#2pt
         \hbox{\vrule width#2pt height#1pt \kern#1pt
               \vrule width#2pt}
               \hrule height#2pt}

\def\Fund#1#2{\vcenter{\vbox{\drawbox{#1}{#2}}}}
\def\Asym#1#2{\vcenter{\vbox{\drawbox{#1}{#2}
               \kern-#2pt       
               \drawbox{#1}{#2}}}}

\def\fund{\Fund{6.4}{0.3}}

\def\bdot{\huge{\textbf{.}}}

\numberwithin{equation}{section}

\begin{document}
\begin{titlepage} 
\begin{flushright}
DCPT-18/29
\vskip -1cm
\end{flushright}
\vskip 4cm
\begin{center}
\font\titlerm=cmr10 scaled\magstep4
    \font\titlei=cmmi10 scaled\magstep4
    \font\titleis=cmmi7 scaled\magstep4
    \centerline{\LARGE \titlerm 
      QCD$_3$ with Two-Index Quarks, Mirror Symmetry}
      \vskip 0.3cm
    \centerline{\LARGE \titlerm and Fivebrane anti-BIons near Orientifolds} 
    \vskip 0.3cm
\vskip 1cm
{Adi Armoni$^\star$ and Vasilis Niarchos$^\natural$}\\
\vskip 0.5cm
       {\it $^\star$Department of Physics, College of Science}\\
       {\it Swansea University, SA2 8PP, UK}\\
\medskip
{\it $^\natural$Department of Mathematical Sciences and Center for Particle Theory}\\
{\it Durham University, Durham, DH1 3LE, UK}\\
\medskip
\vskip 0.5cm
{$^\star$a.armoni@swansea.ac.uk, $^\natural$vasileios.niarchos@durham.ac.uk}\\

\end{center}
\vskip .5cm
\centerline{\bf Abstract}

\baselineskip 20pt
%

\vskip .5cm 
\noindent
We consider a non-supersymmetric Hanany-Witten type IIB brane configuration that realises a three dimensional $USp(2N)$ gauge theory with quarks in the two-index antisymmetric representation and $SO(4)$ flavour symmetry. Using type IIB S-duality we find the mirror dual, an $SO(2N-1)$ field theory with scalars in the antisymmetric representation. Analysing the magnetic dual we study the vacuum structure of the $USp(2N)$ model and propose that the $SO(4)$ global symmetry is unbroken. In order to support our proposal we present an $SO(4)$ symmetric BIon configuration that describes anti-D3 branes polarising into fivebranes in the S-dual Hanany-Witten setup. We also comment on dynamical flavour symmetry breaking in other QCD$_3$ theories with quarks in two-index representations.
\vfill
\noindent
\end{titlepage}\vfill\eject

\setcounter{equation}{0}

\pagestyle{empty}
\small
\vspace*{-0.7cm}

\normalsize
\pagestyle{plain}
\setcounter{page}{1}
 
\setcounter{tocdepth}{2} 
\toc

\section{Introduction} 
\label{intro}

Recently, there has been significant progress in the understanding of the vacuum structure of QCD$_3$ with quarks in the fundamental representation\cite{Appelquist:1988sr,Appelquist:1989tc,Komargodski:2017keh}. There is a lot of evidence that when the number of flavours $N_f$ is within a certain regime that depends on the Chern-Simons level $k$, $k<N_f<N^\star$, the quarks condense and the global flavour symmetry is broken, $SU(2N_f)\rightarrow SU(N_f+k)\times SU(N_f-k)$\cite{Komargodski:2017keh}. On the other hand, little is known about the dynamics of multi flavor QCD$_3$ with quarks in two-index representations, even in the absence of a Chern-Simons term in the action. Our main motivation is to examine more closely the possibility of global symmetry breaking scenarios in such cases. Interestingly, it has been argued recently in a related {\it four dimensional} setup with two adjoint fermions that the global $SU(2)$ flavour symmetry may not be broken \cite{Cordova:2018acb}.

In gauge theories with conserved parity symmetry at the classical level, the Vafa-Witten theorem excludes spontaneous breaking of parity\cite{Vafa:1984xg}. It therefore restricts the possible patterns of symmetry breaking. In particular, in QCD$_3$, when the global symmetry is $SU(2N_f)$ there is either breaking to $SU(N_f)\times SU(N_f)$ or there is no breaking. Similarly, when the global symmetry is $SO(2N_f)$ the only possible pattern is $SO(2N_f)\rightarrow SO(N_f)\times SO(N_f)$. 

In this note we address the issue of dynamical symmetry breaking in a very special setup of three-dimensional gauge theories with fermions in two-index representations, where a string theory embedding can be used to motivate predictions even without supersymmetry.

The starting point of our discussion is a non-supersymmetric Hanany-Witten brane configuration, which contains an orientifold three-plane and anti D3 branes. The field theoretic modes on the D3 branes constitute a non-supersymmetric four dimensional gauge theory on $R^{1,2} \times I$, where $I$ is an interval. When the interval size is taken to zero we obtain a three-dimensional gauge theory with symplectic gauge group, $USp(2N)$. There are four Weyl fermions (quarks) that transform in the two-index antisymmetric representation of the gauge group and the global symmetry is $SO(4)$. We present evidence that this global symmetry does not break dynamically.

Our analysis relies heavily on the dynamics of a non-supersymmetric mirror dual theory, whose existence is motivated by invoking S-duality in type IIB string theory. The dual field theory is constructed using the S-dual Hanany-Witten setup. We explore the dynamics of this setup and propose a scenario where the global $SO(4)\sim SO(3)\times SO(3)$ is not broken. In this scenario quarks may confine but there is no dynamical symmetry breaking.

As further partial evidence in favour of this scenario we construct a D3-D5 BIon solution that describes the mirror dual brane configuration in the limit of large separation between the fivebranes and large three-brane charge. 
We present three different types of non-supersymmetric BIon solutions in the asymptotic background of an $O3^-$ plane. 

The first type is a semi-infinite D5 brane funnel that describes orthogonal anti D3 branes polarised on an ${\bf RP^2}$ in the presence of the $O3^-$. Unlike the usual D3-D5 funnel solution in flat space, the asymptotic size of the ${\bf RP^2}$ is non-zero in the solutions that we discuss and grows with increasing three-brane charge. The solution describes anti D3 branes stabilised at a finite distance away from the orientifold. As a slight generalisation, we demonstrate the existence of such non-supersymmetric BIon solutions in general backgrounds of type IIB supergravity that exhibit repelling forces with anti D3 branes at large radial distances. 

The second type of D5 brane funnel terminates at a finite distance, and like other familiar BIon solutions of the original F1-Dp system \cite{Callan:1997kz}, it can be glued to a mirror to construct a funnel interpolating between a D5 and an anti D5 brane. There is a third type of BIon solutions that terminates on a kink that enters deeps inside the geometry of the orientifold. In this case, we have much less technical control over the solution. We speculate that this type of solutions could be the basis for the construction of BIons interpolating between a pair of largely separated pair of D5 branes. 

All the BIon solutions that we construct preserve the $SO(4)$ symmetry. We discuss the possibility of alternative, symmetry-breaking, BIon solutions and the limitations of the approach.

Although we use a specific setup with $SO(4)$ global symmetry, in the large-$N$ limit there is an equivalence between $USp(2N)$, $SO(2N)$ and $SU(N)$ gauge theories, often called planar equivalence \cite{Armoni:2004uu}. There is also an equivalence between the adjoint/symmetric and antisymmetric representations. We therefore anticipate that in QCD$_3$ theories with quarks in two-index representations there is no dynamical breaking of flavour symmetry.

\section{Mirror Symmetry} 
\label{mirror}

\begin{figure}[!th]
\centerline{\includegraphics[width=6cm]{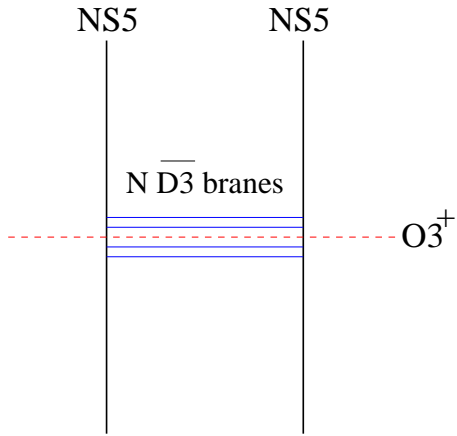}}
\caption{\footnotesize Brane configuration realising the electric theory. It is a non-supersymmetric $USp(2N)$ gauge theory with antisymmetric quarks (and symmetric scalars).}
\label{branes}
\end{figure}

Consider a Hanany-Witten brane configuration \cite{Hanany:1996ie}, which consists of two parallel NS5 branes, whose worldvolume is along the $012345$ directions, an $O3^+$ orientifold plane\footnote{The orientifold changes sign as it crosses an NS5 brane.}, whose worldvolume is along the $0126$ directions and $N$ anti D3 branes and their mirrors, whose worldvolume is along the $0126$ directions. The anti D3 branes are suspended between the two NS5 branes in the 6-direction. The mutual presence of the orientifold and the anti branes breaks supersymmetry. The brane configuration is depicted in figure \eqref{branes}.

The classical field theory on the brane is a $USp(2N)$ gauge theory with 3 real scalars in the symmetric representation and 4 Weyl fermions in the antisymmetric representation. We refer to this theory as the ``electric theory''. This gauge theory may be obtained by dimensional reduction of six dimensional $USp(2N)$ QCD with antisymmetric quarks. The global symmetry (similar to the R-symmetry in the corresponding supersymmetric theory) is $SU(2)_V \times SU(2)_H$. It is realised as $SO(3)_V\times SO(3)_H$ and it is associated with rotations along the directions $345$ and $789$ respectively. The matter content and charges of the gauge theory are listed in table \eqref{table1}.

\begin{table}[!th]
\begin{center}
\begin{tabular}{|c|c|c|c|}
\hline
\multicolumn{4}{|c|} {Electric Theory} \\ 
\hline \hline
 & $USp(2N)$ & $SU(2)_V$ & $SU(2)_H$  \\
\hline
 $A_\mu$ & \Ysymm & \bdot & \bdot  \\
\hline
 $\Phi $ & \Ysymm &  \Ysymm & \bdot  \\
\hline
$\Psi$ & $\T\Yasymm\B$  &  $\fund$ & $\fund$ \\
\hline
\end{tabular}
\caption{\footnotesize The matter content of the electric theory.}
\label{table1}
\end{center}
\end{table}

\begin{figure}[!th]
\centerline{\includegraphics[width=6cm]{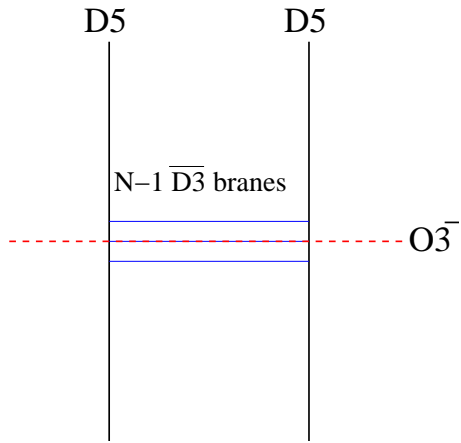}}
\caption{\footnotesize Brane configuration realising the magnetic theory. It is a non-supersymmetric $SO(2N-1)$ theory with antisymmetric scalars (and symmetric fermions).}
\label{branes2}
\end{figure}

Type IIB string theory is believed to admit an exact S-duality. Let us apply it on each component of the brane configuration. The anti D3 branes are invariant under the duality. The NS5 branes become D5 branes. The $O3^+$ plane is transformed into an $\widetilde{O3}^-$ plane, which is morally a bound state of an $O3^-$ plane and a D3 brane. As a result, we expect a tachyonic mode between anti D3 branes when they overlap with the $\widetilde{O3}^-$ plane. Similar to \cite{Uranga:1999ib,Sugimoto:2012rt} we anticipate that the condensation of this mode will lead to $N-1$ anti D3 branes (and their mirror) plus an additional anti D3 brane stuck on top of the $O3^-$ plane. The brane configuration is depicted in figure \eqref{branes2}.

The resulting ``magnetic'' theory at low energies admits a classical global $SO(2N-1)$ symmetry\footnote{Viewed as a four dimensional theory on an interval, the $SO(2N-1)$ symmetry is a gauge symmetry, with heavy gauge bosons.}. It contains 4 real scalars and 4 Weyl fermions. The scalars transform in the antisymmetric representation of $SO(2N-1)$ while the fermions transform in the symmetric representation of $SO(2N-1)$. The low energy matter content of this theory, that follows from classical open string theory, is listed in table \eqref{table2}. $\sigma$ is a real boson and $\phi$ a triplet of real bosons.

\begin{table}[!t]
\begin{center}
\begin{tabular}{|c|c|c|c|}
\hline
\multicolumn{4}{|c|} {Magnetic Theory} \\ 
\hline \hline
 & $SO(2N_c-1)$ & $SU(2)_V$ & $SU(2)_H$  \\
\hline
$ \sigma $ & $\T\Yasymm\B$ &  \bdot & \bdot  \\
\hline
 $\phi $ & $\T\Yasymm\B$ &  \Ysymm & \bdot  \\
\hline
$\psi$ &  \Ysymm &  $\fund$ & $\fund$ \\
\hline
\end{tabular}
\caption{\footnotesize The matter content of the magnetic theory.}
\label{table2}
\end{center}
\end{table}

As we explained, we may view both the electric and magnetic theories as four dimensional gauge theories, formulated on an interval. In the limit where the interval's size goes to zero we obtain the theories listed in \eqref{table1} and \eqref{table2}. From a four dimensional point of view there is a global triangle anomaly that has to be matched. On the electric side the anti-symmetric fermions contribute $4\times 2N (2N-1)$ and on the magnetic side the symmetric fermions contribute $4\times (2N-1) \times (2N-1 +1)$.

We propose that the electric and magnetic theories form a non-supersymmetric dual pair. In fact, in the 't Hooft large-$N$ limit (with the UV cut-off kept finite), the electric and the magnetic theories become supersymmetric, due to planar equivalence \cite{Armoni:2004uu}, and the above duality is the Intriligator-Seiberg mirror symmetry \cite{Intriligator:1996ex}. A scenario for the finite $N$ dynamics of the magnetic theory in the next section will lead to a corresponding conjecture for the vacuum structure of the finite $N$ electric theory.

\section{Dynamics of the electric and magnetic pair} 
\label{dynamics}

Let us start with the electric theory. The gauge theory is non-supersymmetric. The scalars acquire a Coleman-Weinberg potential. As in the corresponding 4d theory \cite{Sugimoto:2012rt}, the generated mass$^2$ of the scalars is positive. In order to see that let us calculate the one-loop contribution to the generated mass (a very similar calculation was carried out recently in the context of a non-supersymmetric 3d Seiberg duality \cite{Armoni:2017jkl}). Both the bosonic and fermionic contributions are proportional to the quadratic Casimir of the corresponding representation. The bosonic contribution is positive while the fermionic contribution is negative. The total contribution is
\beq
M^2 _{\Phi} = g_e^2 \Lambda (2N +2) - g_e^2 \Lambda (2N-2)  >0 \, ,
\eeq
with $g_e$ the electric gauge coupling and $\Lambda$ the UV cut-off. In the string theory context it is natural to identify $\Lambda ^2$ with ${1\over \alpha '}$. The same information can be extracted from the M${\rm \ddot{o}}$bius amplitude which represents the interaction between the anti D3 branes and the orientifold plane. The attraction potential between the branes and the orientifold plane is identified with the attractive Coleman-Weinberg potential for the scalars.

When the field theory limit $\alpha ' \rightarrow 0$ is taken the scalars become infinitely massive and decouple. The low energy theory is simply a $USp(2N)$ QCD$_3$ theory with 4 antisymmetric quarks.

The 3d gauge theory is expected to confine. The main purpose of this note is to address the issue of the vacuum structure and in particular to ask whether the quarks condense. We will focus on $N>1$ \footnote{The case $N=1$ is trivial. For $USp(2)\sim SU(2)$ the antisymmetric quarks decouple from the dynamics.}.

Let us consider the quark bilinear  $\langle \Psi _{\alpha \dot \alpha} \Psi _{\beta \dot \beta} \rangle$. The Lorentz $SO(2,1)$ indices are contracted by using the epsilon symbol and are omitted for brevity. $\alpha, \beta$ are $SU(2)_V$ indices and $\dot \alpha, \dot \beta$ are $SU(2)_H$ indices. A priori, we find three possibilities

\begin{itemize}
\item[$(i)$] $\langle \Psi _{\alpha \dot \alpha} \Psi _{\beta \dot \beta} \rangle = v \delta _ {\alpha \beta} \delta _{\dot \alpha \dot \beta}$. Such a condensate breaks $SU(2)_V\times SU(2)_H \rightarrow SO(2)_V \times SO(2)_H$.

\item[$(ii)$] $\langle \Psi _{\alpha \dot \alpha} \Psi _{\beta \dot \beta} \rangle = \mu \delta _ {\alpha \dot \beta} \delta _{\dot \alpha \beta}$. Such a condensate breaks $SU(2)_V\times SU(2)_H \rightarrow SU(2)_D$.

\item[$(iii)$] $\langle \Psi _{\alpha \dot \alpha} \Psi _{\beta \dot \beta} \rangle = 0$. The global $SU(2)_V \times SU(2)_H$ is unbroken.
\end{itemize}

Now let us consider the dynamics of the magnetic dual (the mirror theory). On the magnetic side the scalars acquire a negative mass$^2$, namely a tachyonic mass, due to the self coupling to antisymmetric scalars and the coupling to symmetric fermions \cite{Armoni:2017jkl}
\beq
M^2 _{\sigma,\phi} = g_m^2 \Lambda ((2N-1) -2) - g_m^2 \Lambda ((2N-1)+2)  <0 \, .
\eeq
As a result, the scalars are expected to develop a vacuum expectation value (vev). The manifestation of the tachyonic mass in the brane picture is a repulsion between the anti D3 branes and the orientifold $O3^-$ plane. This suggests that the $N-1$ anti D3 branes move away from the orientifold (we will study this phenomenon in more detail in the next section). If this happens, the scalars $\phi$ of the magnetic theory acquire a vev and the fermions naturally become massive, due to Yukawa couplings, and decouple. As a result, the low energy dynamics is dictated by the behaviour of scalars. 

Apriori, there are several possibilities. In all of them the $SO(3)_H$ symmetry, associated with rotations along the directions $789$, is unbroken:

\begin{itemize}
\item[$(a)$] The branes attract each other, move away together and sit at a minimum at a finite distance away from the orientifold plane. In this scenario $SO(3)_V\times SO(3)_H\rightarrow SO(2)_V \times SO(3)_H$.

\item[$(b)$] The branes move away and repel each other. They spread away from the orientifold. $SO(3)_V\times SO(3)_H \rightarrow SO(3)_H$.

\item[$(c)$] The branes form a fuzzy sphere. In that case $SO(3)_V\times SO(3)_H$ remains unbroken.
\end{itemize}

Options $(a)$ and $(b)$ are not compatible with the expected dynamics of the electric theory. In scenarios $(i)$ and $(ii)$ on the electric side, where a breaking of the global symmetry occurs, $SO(3)_H$ is broken. The only logically accepted option seems to be option $(c)$ on the magnetic side, which corresponds to option $(iii)$ on the electric side, namely that there is no breaking of the global $SO(3)_V \times SO(3)_H$ symmetry. This implies that there is no quark condensate on the electric side.

Let us elaborate a little further on scenario $(c)$. We follow the analogous 4d scenario described in \cite{Sugimoto:2012rt}, where it was proposed that the branes are distributed in such a way that they preserve $SO(4) \subset SO(6)$. In the present 3d setup the global symmetry is $SO(4)$ and the same distribution of branes does not break the symmetry further. We propose that due to the repulsion between the orientifold and the anti D3 branes, the latter form a configuration of spherical anti D5 branes. We will attempt to find evidence in favour of this scenario with an independent analysis in the next section.

In terms of the scalar expectation values we propose, similarly to \cite{Sugimoto:2012rt}, that the scalars of the magnetic theory $ \{ \phi ^i, \sigma \}$ admit a vev
\begin{eqnarray}
\label{fuzzy}
& & \langle \phi ^i \rangle = c J^i_{(N-1)} \, , \\
& &  \langle \sigma \rangle = c' \, ,
\end{eqnarray}
where $J^i_{(N-1)}$ are generators of the $SU(2)$ algebra with spin $N-1$. Such a configuration preserves the full $SO(3)\times SO(3)$ symmetry. Note that
$c,c'$ are not constants, but a function of the distance between the fivebranes. We omit this dependence in the limit when the interval size is short. In the next section, we describe what happens when the interval is long and a BIon configuration is formed. Evidence in favour of an $SO(4)$ symmetric D3-D5 BIon configuration is presented.

Since we do not have complete control over the dynamics of the magnetic theory we cannot prove rigorously the above scenario. The fact that there exists a configuration, which is consistent with mirror symmetry, is reassuring.

\section{D3-D5 anti-BIons in orientifold backgrounds and generalisations}
\label{bion}

In this section we perform a more detailed study of the mirror brane setup in Fig.\ \ref{branes2} from the perspective of the D5 branes. We analyse D3-D5 BIon solutions that describe semi-infinite polarised anti-D3 branes ending on D5 branes as well as configurations, which are proposed to describe anti-D3s suspended between a $D5$-$\overline{D5}$ pair, or a $D5$-$D5$ pair. The analysis of the D3-D5 BIon solutions that we perform in this section is based on the classical DBI action of the D5 brane on the supergravity background of an $O3^-$ plane. This analysis is expected to be valid in the asymptotic flat space region far away from the orientifold source. We will show that the BIon configurations stabilise in this region when the induced anti-D3-brane charge is sufficiently large. 

There are several approximations that validate this analysis. In the asymptotic region the gradients of the supergravity fields are small and for the most part so are the gradients of the DBI fields. Small gradients are required for the validity of the effective actions that we use (supergravity and DBI). In addition, in the asymptotic region we expect that the backreaction effects of the D5 branes to the supergravity background are subleading and will not alter the qualitative features of the solutions we will find. In particular, we could have performed the same analysis for a large number of overlapping D5 branes in the supergravity regime treating the D5 branes themselves as supergravity solutions. Such a treatment is possible in a regime of long-wavelength approximations using blackfold methods \cite{Emparan:2009cs,Emparan:2009at,Grignani:2010xm,Emparan:2011hg,Niarchos:2012pn,Armas:2016mes}. In this manner, we would indeed find that the leading effects are captured by precisely the same configurations as the one we find with the abelian D5 DBI approach in the present work.  

When contrasted with a treatment based on the effective action of the anti D3 branes,\footnote{A direct treatment from the viewpoint of the anti D3 branes should produce fuzzy brane solutions of the form \eqref{fuzzy}. In this approach, it is not obvious how to determine the effective interactions of the scalars.} one of the immediate benefits of the classical D5 brane perspective is that it incorporates automatically the leading non-supersymmetric one-loop quantum effects on the D3 branes. Some of these effects manifest themselves via the forces that are exerted on the D3-D5 BIon by the background orientifold.

\subsection{A class of supergravity backgrounds}
\label{sugra}

As we explained in section \ref{mirror} after S-duality the $O3^+$-plane, which stretches between a pair of NS5 branes, converts into a $\widetilde{O3}^-$ plane between a pair of D5 branes. A crucial step of the construction of section \ref{mirror} involves the condensation of a tachyonic instability for a single anti D3 brane that overlaps with the $\widetilde{O3}^-$ plane. The condensation of this instability converts this plane to an orientifold that is morally a bound state of an $O3^-$ plane with a half anti D3. 

In subsection \ref{bionSol} we will argue that the half anti D3 brane becomes magnetic flux on the D5 brane. As a result, we have to consider D5 brane configurations in the background of the $O3^-$ plane. In itself the $O3^-$ plane is a supersymmetric configuration. However, in this section we would like to make our analysis a little more general and consider instead a wider class of supergravity backgrounds, which may or may not be supersymmetric. To preserve the obvious symmetries of interest we will consider solutions of the type IIB supergravity equations of motion with a metric of the form (in the Einstein frame)
\beq
\label{sugraaa}
ds^2 = H_1(r) \eta_{ab} dx^a dx^b + H_2(r) (dr^2 + r^2 ds^2_{\bf RP^5})~, ~~ a,b=0,1,2,6
~,
\eeq 
a self-dual RR 5-form field strength 
\beq
\label{sugraab}
F_5 = Q(1+*){\rm Vol}({\bf RP^5})
~,
\eeq
and a potentially non-trivial dilaton
\beq
\label{sugraac}
e^{\phi} = H_3(r)
~.\eeq
$Q$ is a constant (the RR charge of the solution). The functions $H_1, H_2, H_3$ in this ansatz are all functions that depend solely on the radial coordinate $r$.

Interestingly, this general type of solutions has been considered in some generality in the past (mainly with the transverse $\bf RP^5$ replaced by $S^5$, but that does not affect the equations of motion and the corresponding form of the solutions). We refer the reader to \cite{Zhou:1999nm,Lu:2004ms,Nayek:2015tta} for references and \cite{Nayek:2016hsi,Bhattacharya:2018oeq} for some more recent discussions.

Our approach will be the following. Our primary purpose is to establish the existence of non-supersymmetric BIon solutions near the asymptotic flat space region with sufficiently large induced D3 brane charge. Hence, we will focus mainly on the flat space asymptotics of solutions of the form \eqref{sugraaa}-\eqref{sugraac}. The solutions of \cite{Zhou:1999nm} are well suited to this purpose. They are described by a number of free parameters controlling the mass, dilaton charge and RR charge, and have the most general monopole asymptotics of the above-mentioned supergravity fields. There is only one constraint that we will put on these parameters, which will be considered in the next subsection. 

The specifics of the supergravity solutions that we will consider are the following. In the Einstein frame \cite{Zhou:1999nm,Nayek:2015tta}
\beq
\label{sugraad}
ds_E^2 = F^{-1/2}(r) \eta_{ab} dx^a dx^b + F^{1/2}(r) \left( H_+(r) H_-(r) \right)^{1/2} (dr^2 + r^2 ds^2_{\bf RP^5})
~,
\eeq 
\beq
\label{sugraae}
F_5 = Q(1+*){\rm Vol}({\bf RP^5})
~,\eeq
\beq
\label{sugraag}
e^\phi = \left( \frac{H_+}{H_-} \right)^\delta
~.
\eeq
The functions that appear in these expressions are
\beq
\label{sugraai}
F(r) = \left( \frac{H_+}{H_-} \right)^\alpha \cosh^2\omega - \left( \frac{H_+}{H_-} \right)^{-\alpha} \sinh^2\omega
~,
\eeq
\beq
\label{sugraaj}
H_\pm (r) = 1 \pm \frac{R^4}{r^4}
~.
\eeq
The 5-form field strength includes the components $F_{0126r} = Q F^{-2} (H_+ H_-)^{-1} r^{-5}$, which follow from the components of the 4-form RR potential
\bea
\label{sugraaf}
C_{0126} = \frac{Q}{16R^4 \alpha \cosh^4 \omega} \frac{1}{\left(\frac{r^4+R^4}{r^4-R^4} \right)^{2\alpha} - \tanh^2\omega} - \frac{Q}{16R^4 \alpha \cosh^2 \omega}
~.
\eea

The solutions are parametrised by the constant parameters $Q,\alpha, R,\delta,\omega$, but not all of them are independent. The supergravity equations of motion force the following relations between them 
\beq
\label{sugraak}
Q= 8 R^4 \alpha \sinh(2\omega)~, ~~
\alpha^2 + \delta^2 = \frac{5}{2}
~.
\eeq
That leaves three independent parameters, say $\alpha$, $\omega$, $R$. As we see in the next paragraph each of them controls the asymptotics of the three non-trivial fields in the game: the metric, the dilaton and the 4-four RR potential.

Before we proceed further, a few comments about this general solution are in order. As far as we know there is no clean interpretation of these solutions in terms of D3 branes (or $O3$ planes in our context). It is interesting that $H_-$ introduces a singularity at $r=R$. One of the implications of this singularity is that it obstructs the application of Birkhoff's theorem. For our purposes we are not particularly interested in the physics of the singularity, because we only care about the asymptotic large-$r$ region. Moreover, for a solution of small RR charge (as is our case) the curvature (and potentially the dilaton) increase as $r$ become smaller and the usual classical supergravity approximations are invalidated.  

In the asymptotic large-$r$ region the above solutions behave as follows
\beq
\label{sugraal}
ds^2_E \simeq \left( 1- \frac{\alpha R^4 \cosh(2\omega)}{r^4}\right) \eta_{ab}dx^a dx^b + 
\left( 1+ \frac{\alpha R^4 \cosh(2\omega)}{r^4}\right) (dr^2 + r^2 ds^2_{\bf RP^5}) +{\mathcal O}(r^{-8})
~,
\eeq
\beq
\label{sugraam}
C_{0126} \simeq - \frac{Q}{4 r^4} +{\mathcal O}(r^{-8})
~,
\eeq
\beq
\label{sugraan}
\phi(r) \simeq \frac{2 \delta R^4}{r^4} + {\mathcal O}(r^{-12})
~.
\eeq

We note that the supersymmetric solution of the $O3^-$ plane is recovered in the limit $R\to 0$, $|\omega| \to \infty$ with $R^4 \cosh(2\omega)$ kept fixed. It is apparent from the expressions \eqref{sugraai}, \eqref{sugraaj} that in this limit the precise values of $\alpha$ and $\delta$ do not matter and that the solution is expressed in terms of the parameter $Q$. The only information in $\alpha$ that plays a r\^ole is its sign. More specifically, in units where the D3-brane charge is 1, the $O3^-$ charge is $-\frac{1}{4}$. Hence, in these units the parameter $Q$ is a positive number of order 1 (see eq.\ \eqref{sugraam}). Since $Q$ is proportional to $\alpha \sinh(2\omega)$, it is positive when $\alpha \, \omega>0$. The asymptotic form of the metric \eqref{sugraal} and the antigravitating nature of the $O3^-$ plane implies that $\alpha<0$ and therefore also $\omega<0$. We will see that these signs are consistent with the results of the next subsection.

The DBI computation requires the metric in the string frame, $ds^2_S = e^{\phi/2} ds^2_E$, which is listed here for completeness
\beq
\label{sugraao}
ds_s^2 = F^{-1/2}(r) \left( \frac{H_+}{H_-} \right)^{\delta/2} \eta_{ab} dx^a dx^b 
+ F^{1/2}(r) \left( H_+(r) H_-(r) \right)^{1/2} \left( \frac{H_+}{H_-} \right)^{\delta/2} (dr^2 + r^2 ds^2_{\bf RP^5})
~.
\eeq

\subsection{Implications of asymptotic repulsion on anti D3 branes}
\label{D3}

We know that far away from the $O3^-$ plane the anti D3 branes are repelled. In this section we examine what conditions are imposed on the parameters $\alpha$, $\omega$, $R$ by this requirement.

We consider the DBI action for an anti D3 brane in the background metric \eqref{sugraao}, the background dilaton
\eqref{sugraag} and the background RR 5-form flux \eqref{sugraae} part of whose components are
\beq
\label{repulseaa}
F_{0126r} = \partial_r C_{0126} = Q F^{-2}(H_+ H_-)^{-1} r^{-5} 
\eeq
with large-$r$ asymptotics \eqref{sugraam}. The D3-brane is placed parallel to the orientifold source at the transverse position $y^1\equiv y$, $y^2=y^3=y^4=y^5=y^6=0$, i.e.\ at 
$$
r\equiv \sqrt{(y^1)^2 + (y^2)^2 + (y^3)^2 + (y^4)^2 + (y^5)^2 + (y^6)^2}=y
~.
$$
We allow $y=y(t)$ to be a function of the worldvolume time. Then, the induced metric is
\beq
\label{repulseab}
\gamma_{ab} d\sigma^a d\sigma^b = \left( \frac{H_+}{H_-} \right)^{\delta/2} \bigg[ \left(- F^{-1/2} + F^{1/2} (H_+ H_-)^{1/2} (y')^2 \right) dt^2 + F^{-1/2} \left( (dx^1)^2 + (dx^2)^2 + (dx^6)^2 \right) \bigg]
,
\eeq
where the prime in this subsection denotes a derivative with respect to time. The DBI action for the anti D3 brane reads\footnote{We are ignoring the overall tension factor in these expressions for the DBI action. The fact that we have an anti-brane is reflected in the sign in front of the $C_{0123}$ coupling which is $-$ instead of a $+$ for D3s.}
\bea
\label{repulseac}
S_{DBI}&=& - \int dt \, \bigg[ e^{-\phi} \sqrt{-\gamma} + C_{0126} \bigg]
\nonumber\\
&=& - \int dt\, \bigg[ F^{-1} \sqrt{1- F(H_+ H_-)^{1/2} (y')^2 } + C_{0126} \bigg]
\\
&=& -\int dt\, \bigg[ - \frac{1}{2} (H_+ H_-)^{1/2} (y')^2 + F^{-1} + C_{0126} + {\mathcal O}((y')^4) \bigg]
~.\nonumber
\eea
In the asymptotic region, $r=y\to \infty$,
\bea
\label{repulsead}
S_{DBI} &=& -\int dt\, \bigg[ -\frac{1}{2}(y')^2 + 1 - \frac{1}{4} \frac{8R^4 \alpha \cosh(2\omega) + Q}{y^4} +{\mathcal O}(y^{-8}) + {\mathcal O}((y')^4) \bigg]
\nonumber\\
&=& -\int dt\, \bigg[ -\frac{1}{2}(y')^2 + 1 - \frac{Q}{4} \frac{\coth(2\omega) + 1}{y^4} +{\mathcal O}(y^{-8}) + {\mathcal O}((y')^4) \bigg]
~.
\eea
The leading potential for the transverse scalar $y$ in the asymptotic region is therefore
\beq
\label{repulseae}
V(y) \simeq 1 - \frac{Q}{4} \frac{\coth(2\omega) + 1}{y^4}
~.
\eeq
Since we are interested in backgrounds with negative RR charge, eq.\ \eqref{sugraam} implies that $Q>0$, hence there is attraction if $\coth(2\omega) + 1 > 0$, i.e.\ $\omega>0$, and repulsion if $\coth(2\omega)+1<0$, i.e. $\omega<0$. We want to capture a situation where the latter happens, hence we take
\beq
\label{repulseaf}
\omega < 0
~.
\eeq
Since $Q=8 R^4 \alpha \sinh(2\omega) >0$ we must also take $\alpha<0$. This is consistent with the negative sign of $\alpha$ that was argued for the $O3^-$ plane in the previous subsection.

\subsection{BIon solutions}
\label{bionSol}

The abelian DBI action for a D5 brane is
\beq
\label{warmaa}
S_{DBI} = - T_5 \int d^6 \sigma\, e^{-\phi} \sqrt{-\det (\gamma_{ab} + B_{ab} + 2\pi \alpha' F_{ab})} +\mu_5 \int d^6 \sigma \, \sum_p C_{(p+1)} \wedge e^{2\pi \alpha' F + B}
~.
\eeq
$\gamma_{ab}$ is the induced metric, $F_{ab}$ the abelian gauge field on the D5 brane, $B_{ab}$ the pull-back of the background Kalb-Ramond field and $C_{(p+1)}$ background $(p+1)$-form potentials. In the case of the supergravity solutions of subsection \ref{sugra}, the only non-vanishing background fields are the dilaton and the four-form $C_4$.

The planar D5 brane is oriented along the spacetime coordinates 012345. Let us parametrize the 345 plane in spherical coordinates as
\beq
\label{warmab}
(dx^3)^2 + (dx^4)^2 + (dx^5)^2 = d\rho^2 + \rho^2 \left( d\psi^2 + \sin^2\psi\, d\theta^2 \right)
~.
\eeq
The coordinates $(\psi,\theta)$ parametrise the ${\bf RP^2}$, hence they obey the discrete identification
\bea
\label{mainac}
&&\psi \leftrightarrow \psi~, ~~ \theta \leftrightarrow \pi-\theta~~~{\rm for}~~ x^3 \to -x^3
\nonumber\\
&&\psi \leftrightarrow \psi~, ~~ \theta \leftrightarrow -\theta~~~~\, \, \, \, {\rm for}~~ x^4 \to -x^4
\nonumber\\
&&\psi \leftrightarrow \pi- \psi~, ~~ \theta \leftrightarrow \theta~~~{\rm for}~~ x^5 \to -x^5
~.
\eea

We want to find spike/funnel solutions that describe D3 branes emerging orthogonally out of the D5 brane worldvolume in the directions 0126. Therefore, we make the following ansatz for the worldvolume coordinates in static gauge
\beq
\label{warmac}
X^0 = \sigma^0 ~, ~~ X^1 = \sigma^1~, ~~ X^2 = \sigma^2~, ~~ \psi = \sigma^3~, ~~ \theta = \sigma^4~, ~~ \rho = \sigma^5
~.
\eeq
We activate only one of the transverse scalars, $X^6$, which will be called for convenience $X$. We set
\beq
\label{warmad}
X = X(\rho)
\eeq
and denote the derivative of $X$ with respect to $\rho$ as $X'$. The D5 brane is positioned at $X^7=X^8=X^9=0$. With these specifications the induced metric is
\bea
\label{bionaa}
\gamma_{ab}d\sigma^a d\sigma^b &=& \left( \frac{H_+}{H_-} \right)^{\delta/2} \bigg[ F^{-1/2} (-(d\sigma^0)^2 + (d\sigma^1)^2 + (d\sigma^2)^2 ) 
\\
&&+F^{1/2} (H_+ H_-)^{1/2} \rho^2 (d\psi^2 + \sin^2\psi d\theta^2) 
+\left( F^{1/2} (H_+ H_-)^{1/2} + F^{-1/2} (X')^2 \right)d\rho^2 \bigg]
~.\nonumber
\eea

A D3 brane charge can be induced by turning on the following gauge field strength on the D5 brane worldvolume
\beq
\label{gaugefield}
F= \frac{g}{2\pi \alpha'} \, \sin\psi \, d\psi \wedge d\theta
~.
\eeq
This is the field strength of a magnetic monopole. The constant $g$ controls the induced D3 brane charge. It is straightforward to check that this ansatz satisfies the DBI equations of motion for the abelian gauge field on the D5 brane.

Since the second Stiefel-Whitney class of ${\bf RP^2}$ is non-trivial, it has been argued \cite{Aharony:2000cw} (see also \cite{Hyakutake:2000mr} for related discussions) that the flux of $\int_{\bf RP^2} F$ is quantized in half-integer units in the background of the $O3^-$ plane. That means that the half anti D3 brane, which is part of the $\widetilde{O3}^-$ plane in Fig.\ (\ref{branes2}) should be treated as flux on the D5 brane solution that we are looking for.\footnote{We would like to thank Shigeki Sugimoto for discussions related to this point.} Henceforth, we assume that the constant $g$ in \eqref{gaugefield} is quantised in this fashion. In addition, the $O3^-$ plane has trivial discrete torsion for the NSNS 3-form field strength, hence we consider no contributions to the DBI action arising from the Kalb-Ramond 2-form field $B_2$.

With the above ansatz we obtain the following effective action for the transverse scalar $X(\rho)$ from the DBI action  
\bea
\label{bionab}
S=-\int d\rho\, \bigg\{
F^{-1} \bigg[ \left( \frac{H_+}{H_-} \right)^\delta H_+ H_- F \rho^4 + g^2 \bigg]^{1/2} 
\sqrt{F(H_+ H_-)^{1/2} + (X')^2}
+g C_{0126}(\rho) X' \bigg\}
~.\nonumber\\
\eea
In this expression we have omitted a trivial overall tension factor and $C_{0126}(\rho)$ is given by the formula \eqref{sugraaf}, which can also be written as 
\beq
\label{bionad}
C_{0126} = \frac{\sinh\theta}{\cosh^3\omega} \frac{1}{\left(\frac{\rho^4+R^4}{\rho^4-R^4} \right)^{2\alpha} - \tanh^2\omega} - \tanh\omega
~.
\eeq
Varying with respect to $X(\rho)$ we obtain the equation of motion
\bea
\label{bionac}
X'(\rho) = \pm \frac{(c-g C_{0126}) F^{1/2} (H_+ H_-)^{1/4}}
{\sqrt{F^{-2}\left( \left( \frac{H_+}{H_-} \right)^{\delta/2} H_+ H_- F \rho^4 + g^2 \right) - (c-gC_{0126})^2}}
~,
\eea
where $c$ is an integration constant.

\subsubsection{Semi-infinite spikes}
\label{spike}

The first type of solution to \eqref{bionab} that we are looking for is a solution that describes a semi-infinite spike that stretches along $X^6 \equiv X \in [0,+\infty)$. For this type of solution we require the boundary conditions
\beq
\label{spikeaa}
\lim_{\rho\to \infty} X(\rho) =0
~.
\eeq
In analogy with the original F1-D$p$ BIons of \cite{Callan:1997kz} the generic solution with this boundary condition will not be a spike stretching across the whole semi-infinite positive $X$-line. It will terminate at some $X$ for a specific $\rho$. What happens depends on the roots of the quantity
\beq
\label{bionae}
{\mathfrak G} \equiv F^{-2}\left( \left( \frac{H_+}{H_-} \right)^{\delta/2} H_+ H_- F \rho^4 + g^2 \right) - (c-gC_{0126})^2
\eeq
under the square root in the denominator of \eqref{bionac}. In cases where $\mathfrak G$ has a simple root the solution will terminate at a finite $\rho_*$ and a finite $X_*$. Near this point the solution behaves as $X\sim X^*+ a \sqrt{\rho-\rho_*}$ for some constant $a$. If there is a double root at $\rho_*$, then the spike will go all the way to infinity where it will behave logarithmically as $X\sim a \log (\rho - \rho_*)$. In this subsection we are exclusively interested in the possibility of this latter behaviour. The question is whether we can tune the integration constant $c$ to obtain a double root of $\mathfrak G$.

We can be more explicit about this question in the asymptotic region, which is also the region of main interest. At large $\rho$
\beq
\label{bionaf}
{\mathfrak G} \simeq \frac{1}{\rho^4} \left( \rho^8 + A \rho^4 + B \right) +{\mathcal O}(\rho^{-8})
~,
\eeq
where
\bea
\label{bionag}
A &=& - c^2 + g^2 + \delta R^4 - 2 \alpha  R^4 \cosh(2\omega)~,
\\
B&=& \frac{R^4}{4}\Big( -8\alpha ( 2g^2 +  \delta  R^4 ) \cosh(2\omega) + R^4 ( 1-2\alpha^2 + 8 \alpha^2 \cosh(4\omega) ) -16 c g \alpha \sinh(2\omega) \Big)
~.\nonumber\\
\eea
Hence, $\mathfrak G$ will exhibit a double root in the asymptotic region when the discriminant of $x^2+Ax+B$ vanishes, i.e.\ when
\beq
\label{bionai}
\Delta \equiv A^2 - 4B =0
~.
\eeq
This is a quartic polynomial equation on the integration constant $c$. We expect to have a solution that terminates at large $\rho$ when $g$ is large (which is verified by the numerics). At large $g$, and all the other background parameters fixed, we find that the discriminant of the quartic equation \eqref{bionai} goes like $4096 R^8 \alpha^2 g^8$, hence it is positive. In the special case of the $O3^-$ plane, where $R\to 0$, $\omega\to -\infty$ with $Q$ fixed, the discriminant goes like $128g^8 Q^3$ in the large $g$ limit, hence it is again positive. As a result, \eqref{bionai} either has four distinct real roots or two pairs of complex-conjugate roots. Let the quartic equation \eqref{bionai} be of the form 
\beq
\label{bionaj}
{\mathfrak a} x^4 + {\mathfrak b} x^3 + {\mathfrak c}x^2 + {\mathfrak d}x +{\mathfrak e}=0
~.
\eeq
To determine which case is realised one should compute the quantities
\beq
\label{bionak}
P \equiv 8 {\mathfrak a}{\mathfrak c}-3 {\mathfrak b}^2~, ~~
D \equiv 64 {\mathfrak a}^3 {\mathfrak e}- 16 {\mathfrak a}^2 {\mathfrak c}^2 + 16 {\mathfrak a} {\mathfrak b}^2 {\mathfrak c} - 16 {\mathfrak a}^2 {\mathfrak b}{\mathfrak d} -3 {\mathfrak b}^4
~.
\eeq
If $P<0$ and $D<0$, then all four roots are real and distinct. If $P>0$ or $D>0$, then all the roots are complex. In the large-$g$ limit we find
\beq
\label{bional}
P \simeq -18 g^2 ~, ~~
D \simeq 256\, \alpha g^2 R^4 \cosh(2\omega)
~.
\eeq
Both of these expressions take a finite negative value in the limit of parameters relevant for the $O3^-$ case. Consequently, we conclude that there are real values of $c$ for which we get semi-infinite BIons with a finite size core when $\alpha<0$. In contrast, there are no solutions of this type for real $c$ when $\alpha>0$. This is precisely what one would naturally anticipate. A semi-infinite BIon with a finite size core exists when the asymptotic force on an anti D3 brane is repulsive! Notice that in flat space the well known semi-infinite D3-D5 BIon is a spike with vanishing $S^2$ radius at infinity. Here the solutions have an ${\bf RP^2}$ with non-vanishing radius at infinity. 

Another noteworthy feature of the above solutions for $\alpha<0$ is that for any given value of $g$ there are two values of $c$ that give a semi-infinite BIon on the negative $X$-axis and two values of $c$ that give a semi-infinite BIon on the positive $X$-axis. We will not attempt to determine which of these solutions is energetically favourable since our main purpose in this paper is to exhibit the existence of these solutions. It would be interesting, however, to examine more closely the stability properties of these solutions. On physical grounds we anticipate that at least one of them is metastable as a result of the competing effects of the repulsion from the orientifold source and the attractive nature of the D5 tension.

\begin{figure}[!t]
\centerline{\includegraphics[width=8cm]{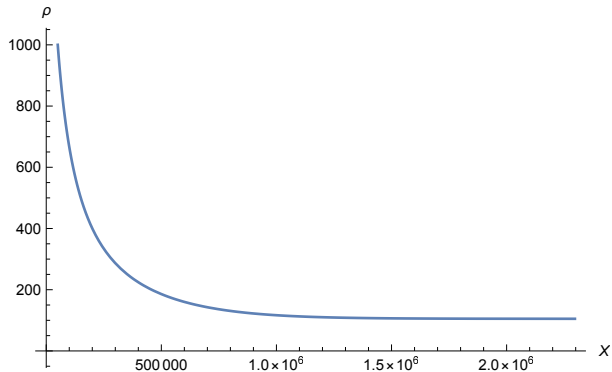}}
\caption{\footnotesize The shape of a semi-infinite BIon solution for the parameters $\alpha=-1$, $\omega=-0.5$, $g=10^8$. The plot depicts the $\bf RP^2$ size of the BIon, $\rho$, as function of the transverse distance $X$. In this particular case we used one of the critical values of $c$, $c\simeq 10^8+1.21306127$. The size of the BIon funnel at $X\to \infty$ is $\rho\simeq 104.95$. To get the numerical solution we used the boundary condition $X(2000)=0$ as an approximation to \eqref{spikeaa}. Very similar solutions exists in the O3$^-$, where $R \rightarrow 0$, $\omega \rightarrow -\infty$ with $R^4 \cosh (2\omega)$ kept fixed.}
\label{plot_spike1}
\end{figure}

We have verified the above statements with explicit numerical computations.  In Fig.\ \ref{plot_spike1} we present the shape of the semi-infinite BIon solution for a particular choice of parameters.

\subsubsection{Solutions describing anti-D3s suspended between D5s}
\label{suspend}

\begin{figure}[!t]
\centerline{\includegraphics[width=9cm]{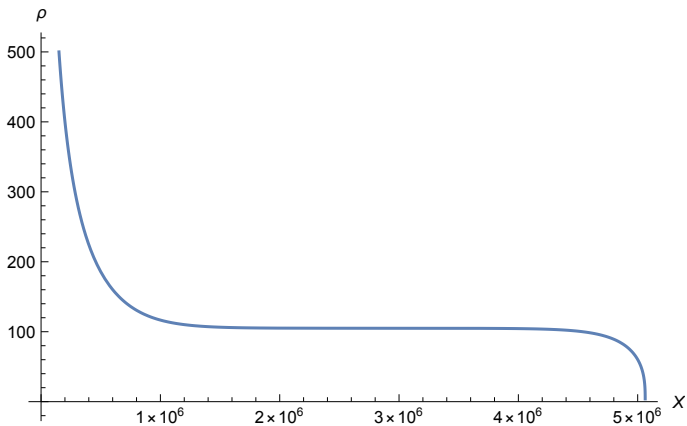}}
\caption{\footnotesize The shape of a semi-infinite BIon solution for the parameters $\alpha=-1$, $\omega=-0.5$, $g=10^8$. The plot depicts the $\bf RP^2$ size of the BIon, $\rho$, as function of the transverse distance $X$. In this particular case we used $c = 10^8+1.21306126$. The size of the BIon funnel reduces drastically at $X\simeq 4.8\times 10^6$. To get the numerical solution we used the boundary condition $X(2000)=0$ as an approximation to \eqref{spikeaa}. Solutions of this type also exist in the O3$^-$ case.}
\label{plot_spike2}
\end{figure}

The solutions of the previous subsection describe semi-infinite anti D3 branes ending orthogonally on a single D5 brane. In this paper we are interested in solutions that describe anti D3 branes suspended between two D5 branes separated by a finite distance. For that purpose we need to examine what happens to the solutions of equation \eqref{bionac} for integration constants $c$ different from the critical values $c_*$ that lead to double roots in $\mathfrak G$. Let us call $\rho_{double}$ the double root at $c_*$.

When we deform $c$ away from $c_*$ the solution exhibits two qualitatively different behaviours. Decreasing $c$ below $c_*$ removes all roots (double or simple) in the vicinity of $\rho_{double}$. In Fig.\ \ref{plot_spike2} we depict what happens to the semi-infinite spike of Fig.\ \ref{plot_spike1} when we shift $c$ from the critical double-root value $c=10^8+1.21306127$ slightly away to $c = 10^8+1.21306126$. For a very large range of $X$ values the solution behaves as a semi-infinite BIon with a finite ${\bf RP^2}$ size, but at some large finite $X$ the size of the ${\bf RP^2}$ decreases dramatically and the D5 funnel moves deeply towards the center of the background geometry. When this happens we lose control of the approximations that were assumed in the beginning. Let us call this type of solutions {\it type I}. We will return to these solutions in a moment.

\begin{figure}[!t]
\centerline{\includegraphics[width=9cm]{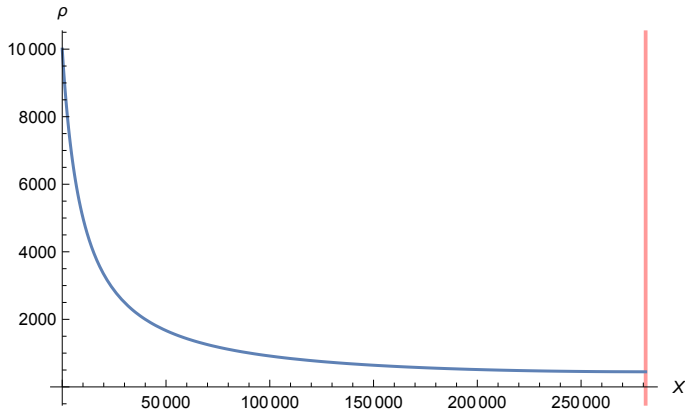}}
\caption{\footnotesize The shape of a semi-infinite BIon solution for the parameters $\alpha=-1$, $\omega=-0.5$, $g=10^8$. The plot depicts the $\bf RP^2$ size of the BIon, $\rho$, as function of the transverse distance $X$. In this particular case we used $c = 10^8+200$. The BIon funnel terminates at $X\simeq 281100$, where $\rho\simeq 447.25$. The red vertical lines denotes the termination point. To get the numerical solution we used the boundary condition $X(10000)=0$ as an approximation to \eqref{spikeaa}. Solutions of this type have been verified also in the O3$^-$ case.}
\label{plot_spike3}
\end{figure}

The second possibility is to increase the value of $c$ above $c_*$. In that case the double root of $\mathfrak G$ breaks up into two simple roots. The solution terminates at a size $\rho$ which coincides with the larger of the two simple roots. As we mentioned previously, the solution behaves near that simple root as $X\sim X^*+ a \sqrt{\rho-\rho_*}$ for some constant $a$. In this manner we obtain a BIon with a finite $\bf RP^2$ size that terminates at a finite distance $X$ away from the D5. We will call this type of solution {\it type II}. Fig.\ \ref{plot_spike3} demonstrates an example. This is similar to the finite throat solutions that one finds in F1-D3 \cite{Callan:1997kz}, or D3-D5 BIon solutions in flat space. 

As in \cite{Callan:1997kz} it is straightforward to glue two mirror type II solutions to obtain a wormhole BIon connecting a D5-anti-D5 pair. In this case there is an abelian $F_2$ flux along a single D5 brane whose strength is proportional to $\frac{g}{2\pi\alpha'}$. This factor is constant throughout the solution.

The construction of a BIon configuration that describes D3s suspended between a D5-D5 pair is less straightforward. One of the issues has to do with the $F_2$ flux. Gluing together two mirror BIon solutions requires that the constant $g$ is positive on the left half of the configuration and negative on the right half. As a result, if the $\bf RP^2$ size of the BIon throat is everywhere non-vanishing, then the $F_2$ flux cannot vanish either and the passage from $g$ to $-g$ along the $X$ direction will be discontinuous. This suggests that a continuous configuration requires a kink in the middle of the interval between the D5s, where the $\bf RP^2$ size of the BIon vanishes and, in a naive sense, the BIon touches the orientifold. We notice that the type I BIon solutions that we found above have exactly this feature. Therefore, it is very tempting to propose that D3-branes suspended between a pair of separated D5 branes are described by gluing together two mirror type I BIon solutions. The parameter $c$ in these solutions would control the separation of the D5 branes. 

The fate of this proposal relies on the details of the configuration close to the orientifold. Unfortunately, this is a region where most of our approximations in this paper break down. This includes both the supergravity approximations and the approximations that lead to the DBI action. As we see for example in Fig.\ \ref{plot_spike2}, in the limit of large separation between the D5 branes the dangerous effects of the kink in the middle of the configuration are concentrated in a small region whose size becomes smaller and smaller as the separation of the D5 branes increases. It is impossible with current techniques to check explicitly what happens in this region and whether one can validate a solution with this behaviour. 

The above solutions are constructed with an ansatz that respects the $SO(3)\times SO(3)$ symmetry. It is interesting to ask whether there are other solutions that violate this symmetry. For example, one can try to solve the DBI equations for a D5 with magnetic flux \eqref{gaugefield} with an ansatz that activates a more general combinaton of the transverse scalars $X^6, X^7, X^8, X^9$. Such an ansatz would describe polarised anti D3 branes stretching between two D5 branes along a curved line inside the 6789 plane. Independent of the details of this curve, one would still have to face the issue of the previous paragraphs. Since the flux on the D5 brane has to change sign at some point, there will be a region where the funnel has to shrink and curve towards the core of the background geometry. We have not established the existence of such solutions, but even if we did, it would be unclear if they would be energetically dominant or subdominant compared to the solutions along $X^6$ that we presented above.

\section{Discussion} 
\label{discussion}

In this short note we used mirror symmetry to argue that the global $SO(4)$ flavour symmetry is not broken in our model, for all values of the gauge group rank $N$. We used S-duality in type IIB string theory to motivate a non-supersymmetric mirror symmetry duality in field theory. We argued that the mere statement of duality puts constraints on potential patterns of spontaneous symmetry breaking. The most plausible scenario seems to be one that favours the lack of spontaneous breaking of the global symmetry. In this scenario the scalars of the dual magnetic theory cannot condense in an arbitrary way if they are to match natural expectations about mirror symmetry and the dynamics of the electric theory.

The existence of $SO(4)$ symmetric D3-D5 BIon configurations in the S-dual brane setup that describe polarised anti D3 branes stabilised at a finite distance away from the orientifold appears to add some evidence in the same direction, but it is not conclusive. The analysis of such solutions in section \ref{bion} was performed in a limit of large induced three-brane charge and large D5 brane separation. The $SO(4)$ symmetric semi-infinite spike solutions with a finite ${\bf RP^2}$ radius in subsection \ref{spike} are on solid ground in these limits and were established for a large class of supergravity backgrounds. They exhibit that a mechanism of stabilisation of polarised anti D3s away from the orientifold exists. Unfortunately, the case of BIon solutions that interpolate between two parallel D5 branes is more subtle. In that case, one cannot avoid a region of space where the validity of our approximations breaks down. The hard technical issues associated with this issue have not allowed us to establish conclusively that anti D3 branes suspended between two parallel D5 branes polarise in an $SO(4)$ symmetry way. A conclusive resolution of this problem would be interesting per se as a result about non-supersymmetric dynamics in string theory.  

Finally, in this paper we addressed the issue of symmetry breaking in a special setup with four Weyl antisymmetric quarks. Let us assume that the lack of flavour symmetry breaking is correct and persists in similar $USp(2N)$ three dimensional gauge theories with $N_f$ antisymmetric quarks. If that case, such a breaking of the flavour symmetry will not occur also in large-$N$ $SO(2N)$ or $SU(N)$ gauge theories with quarks in either the symmetric/antisymmetric or adjoint representations. The reason is planar equivalence \cite{Armoni:2004uu}: all these representations and gauge groups become equivalent in the large-$N$ limit. It is possible that such a breaking does not occur at finite $N$ as well. It would be interesting to explore this possibility in more detail in the future.

The question about the relation between dynamical flavour symmetry breaking and confinement is also fascinating. In four dimensions there exists an argument due to Casher that links chiral symmetry breaking with confinement \cite{Casher:1979vw}. In three dimensions there is no notion of chirality and hence Casher's argument does not apply to 3d. It is therefore possible that in a 3d gauge theory quarks will confine, but will not condense. That may be the case when the quarks are in two-index representations. This is another interesting question for future investigations.

\subsection*{Acknowledgements}

 The work of AA has been supported by STFC grant ST/P00055X/1. We thank Zohar Komargodski and Shigeki Sugimoto for useful discussions and comments on the manuscript. The work of V.N. is supported by STFC under the consolidated grant ST/P000371/1.

\bibliographystyle{utphys}
\providecommand{\href}[2]{#2}

\bibliographystyle{utphys}
\end{document}